\documentclass{article}

\usepackage{PRIMEarxiv}

\usepackage[utf8x]{inputenc} 
\usepackage[T1]{fontenc}    
\usepackage{url}   
\usepackage{bm}
\usepackage{caption}
\usepackage{float}
\usepackage[dvipsnames]{xcolor}
\usepackage{subcaption}
\usepackage{graphicx}
\usepackage{dcolumn}

\usepackage{sectsty}
\usepackage{float}
\usepackage{fancyhdr}
\usepackage{fnpos}
\usepackage[english]{babel}
\usepackage{lastpage}
\usepackage{booktabs}       
\usepackage{amsfonts}
\usepackage{mathptmx}
\usepackage{mathtools}
\usepackage{amsfonts}
\makeatletter
\def\blfootnote{\xdef\@thefnmark{}\@footnotetext}
\makeatother
\usepackage{nicefrac}       
\usepackage{microtype}      
\usepackage{lipsum}
\usepackage{fancyhdr}       
\usepackage{graphicx}       
\graphicspath{{media/}}     

\title{Ising chain: Thermal conductivity and first-principle validation of Fourier law

}

\author{
   \noindent\large{Henrique Santos Lima\textit{$^{a}$}, Constantino Tsallis\textit{$^b$}}}

\begin{document}
\maketitle

\begin{abstract}
The thermal conductivity of a $d=1$  lattice  of ferromagnetically coupled planar rotators is studied through molecular dynamics. Two different types of anisotropies (local and in the coupling) are assumed in the inertial  XY model. 
In the limit of extreme anisotropy, both models approach the Ising model and its thermal conductivity $\kappa$, which, at high temperatures, scales like $\kappa\sim T^{-3}$. This behavior reinforces the result obtained in various $d$-dimensional models, namely  
$\kappa \propto L\, e_{q}^{-B(L^{\gamma}T)^{\eta}}$ where $e_q^z \equiv[1+(1-q)z]^{\frac{1}{1-q}}\;(e_1^z=e^z)$, $L$ being the linear size of the $d$-dimensional macroscopic lattice. The scaling law
$\frac{\eta \,\gamma}{q-1}=1$ guarantees the validity of Fourier's law, $\forall d$.
\end{abstract}

\blfootnote{\textit{$^{b}$~Centro Brasileiro de Pesquisas Fisicas, Rua Xavier Sigaud 150, Rio de Janeiro-RJ 22290-180, Brazil.\\ 
E-mail: hslima94@cbpf.br }}

\blfootnote{\textit{$^{b}$~Centro Brasileiro de Pesquisas Fisicas and National Institute of Science and Technology of Complex Systems, Rua Xavier Sigaud 150, Rio de Janeiro-RJ 22290-180, Brazil \\
Santa Fe Institute, 1399 Hyde Park Road, Santa Fe, 
 New Mexico 87501, USA \\
Complexity Science Hub Vienna, Josefst\"adter Strasse 
 39, 1080 Vienna, Austria \\
E-mail: tsallis@cbpf.br}}

\section{Introduction}
Transport properties naturally emerge in macroscopic systems which are not in thermal equilibrium. For instance, if a system is in permanent contact with two or more reservoirs having different temperatures,  electrical  potentials, concentrations and mean velocities, transfer of heat and of similar quantities (charge, mass, momentum) spontaneously occur. These phenomena lead to linear relations between causal quantities (appropriate gradients, assumed to be asymptotically small) and their effects (corresponding transfers,) yielding  characteristic coefficients such as  thermal conductivity, electrical conductivity, diffusivity and viscosity, appearing respectively in Fourier's, Ohm's, Fick's and viscosity Newton's laws. 

We focus here on Fourier's law. In the absence of radiation and convection,  this law~\cite{Fourier1822} consists in a linear relation between heat flux $\mathbf{J}$ and the gradient of the temperature field $-\nabla T$ which causes this flux,  thus yielding, at the stationary state, the well known relation $\mathbf{J}/L^{d-1}=-\kappa \nabla T$, being $\kappa >0$ referred to as the {\it thermal conductivity} of the $d$-dimensional medium and $L$ its linear size. 

This important transport property currently satisfies some rules, namely that $\kappa$ neither depends on the  gradient of the temperature as long as it is small, nor on the system size as long as it is large~\cite{Raymondetal}. This centennial law was usually  used to tackle with three-dimensional materials, because  in the nineteenth century atoms and molecules were only theoretical particles, with no direct evidence of their existence. Nowadays, various low-dimensional systems have been experimentally \cite{Xu2016,Wu2017} and theoretically \cite{Buttner1984,Laurencot1998,Aoki2000,Gruber2005,Bernardin2005,Bricmont20071,Bricmont20072, Wu2008,Gaspard2008} investigated  and some of them, even in one-dimension, obey that important relation ~\cite{Lebowitz1978, Michel2003,LandiOliveira2014, TsallisLima2023}. For instance, planar rotators following an inertial  $XY$-model obey Fourier's law for all dimensions~\cite{TsallisLima2023}. In contrast, it has also been claimed that this law is violated in cases such as  ballistic diffusion regime~\cite{DubiVentra20091,DubiVentra20092}, non-momentum conserving systems ~\cite{Gerschenfeld2010}, and anomalous heat diffusion~\cite{Dahr2011,LiuXu2012,YangZhang2010}. Let us also mention its possible experimental invalidity in carbon nanotubes~\cite{HanFina2011}.

Paradigmatic ferromagnets are in general described by a set of interacting spins in a crystalline $d$-dimensional lattice that contains $n$ spin vector components such that $|\mathbf{S}|=1$ . In the absence of external fields and inertial terms, the Hamiltonian of these systems can be expressed in the following form:
\begin{equation}
    \mathcal{H}=-J\sum_{\langle ij \rangle}\sum_{m=1}^n  S_i^{m}S_j^{m}\;\;(J>0;\,\sum_{m=1}^n (S_i^{m})^2 =1)\,,
    \label{nvec}
\end{equation}
where $\langle ij \rangle$ denotes first-neighboring spins, and $n=1,2,3,\infty$ correspond respectively to the Ising, $XY$, Heisenberg and spherical models \cite{Stanley1968}.
 Their transport properties have been little  investigated in the literature~\cite{Savin2005,Louis2006, LiLiLi2015, LiLi2017,OlivaresAnteneodo2016,Aziz2022}. In particular, the Ising model has no dynamics, and is therefore unfeasible  by  molecular dynamics approaches. Extensions of Monte Carlo techniques exist~\cite{HarrisGrant1988}, but these methods are not grounded on first-principles and no information about the evolution of the system can be provided. For instance, if the system is in a non-stationary state, the heat flux   fluctuates  and can be  positive or negative. Within a molecular dynamics approach, all information is available at any instant of time. Knowledge about the Ising case (i.e., $n=1$) might be important for the thermal control of spin excitations~\cite{Menaetal2020,Kojimaetal2022} and  skyrmion-hosting materials~\cite{Chauhanetal2022}, among others.

In the present study we approach, for a linear chain, the Ising limit via  two different types of extremely anisotropic $XY$ models, namely through the addition of a local term in the Hamiltonian (preliminary discussed in \cite{Borgesetal2003}), or by allowing the $XY$ interaction to be anisotropic.

\section{Models and methods}
Let us first focus on the local possibility. We assume that the Hamiltonian of the inertial $XY$ model includes a  local energy being proportional to a self-interaction between  spins  in the $x$-direction. This Hamiltonian can then be  written as follows
\begin{equation}
\label{ham}
    \mathcal{H}_{XY}^{\,l}=\sum_{i=1}^L\frac{p_i^2}{2}+\frac{1}{2}\sum_{\langle i,j\rangle}\left[1-\cos(\theta_i-\theta_j)\right]+\epsilon_l\sum_{i=1}^L \sin^2{\theta_i}\,,
\end{equation}
where $\epsilon_l \in [0,\infty]$ is a coupling constant associated with this local energy. This model is similar to Blume-Capel  one~\cite{Blume1966,Capel1966}, but with $n=2$ instead of $n=1$. For  increasing $\epsilon_l$, the second term in the Hamiltonian dominates, thus exhibiting  properties that are characteristic of the $n=1$ class of the Hamiltonian Eq. \eqref{nvec}. The $\epsilon_l\to\infty$ corresponds to a complete crossover from the $XY$ model to the Ising one.


The Hamiltonian described by  Eq.~\eqref{ham}, adding Langevin heat baths acting only on the first and the last particles of the chain with  temperatures $T_h$ and $T_l$ ($T_h\ge T_l$) respectively. The corresponding equations of motion are given by 
\begin{align}
\begin{split}
\label{langevin} 
\dot \theta_i&=p_i\,\,\,\text{for $i=1, \dots, L$}\\
    \dot p_1&=-\gamma_h p_1+F_1+\eta_h(t)\\
    \dot p_i &=F_i \,\,\,\text{for $i=2,\dots, L-1$}\\
    \dot p_L&=-\gamma_l p_L+F_L+\eta_l(t) 
    \end{split}  
\end{align}

where the force components are

\begin{equation}
    F_i=-\epsilon_l\sin(2\theta_i)-\sum_{\langle j\rangle} \sin(\theta_i-\theta_j)\,,  
    \label{force1}
    \end{equation}
where, for each $i$, $\langle j\rangle$ means that we are summing over  nearest-neighbor pairs;  $\eta_{h/l}$ are Gaussian white noises with  correlations 
\begin{align}
\begin{split}
\langle \eta_{h/l}(t)\eta_{h/l}(t')\rangle&=2\gamma_{h/l}T_{h/ l}\delta(t-t')\\ 
\langle \eta_{h}(t)\eta_{l}(t')\rangle&=0\,.
    \end{split}
\end{align}
The heat flux is derived via continuity equation; the Lagrangian heat flux $J_i$  ~\cite{Mejia2019} is given  by
\begin{equation}
J_i=\frac{1}{2}(p_i+p_{i+1})\sin(\theta_i-\theta_{i+1})\,.
\label{flux1}
\end{equation}
Let us emphasize that Eq. \eqref{flux1} has the same form $\forall \epsilon_l$, i.e. that
of the Lagrangian heat flux of the  $XY$ model itself~\cite{ TsallisLima2023}. Despite of the fact that the local term does not contribute to the structure of the heat flux, the evolution of the  canonical coordinates  is quite different for different values of $\epsilon_l$. Indeed,  the presence of the local force [Eq. \eqref{force1}] enters into the average $J\equiv \langle J_i \rangle_{bulk} $, which is then  affected by $\epsilon_l$. The thermal conductance $\sigma$ of the chain is defined as follows 
\begin{equation}
\label{steady}
    \sigma \equiv \frac{\kappa}{L}=\frac{J}{T_h-T_l} .
    \end{equation}
This definition is obtained through the one-dimensional heat equation $\frac{\partial T}{\partial t}\propto\frac{\partial^2 T}{\partial x^2}$. In the steady state,  $\frac{\partial T_{st}}{\partial t}=0$, where $T_{st}= T_{st}(x)$ is the  steady-state temperature field. By imposing the boundary conditions $T_{st}(0)=T_h$ and $T_{st}(L)=T_l$ we have the solution $T_{st}(x)=\frac{T_l-T_h}{L}x+T_h$, hence the heat flux is given by
\begin{equation}
    J=\kappa\frac{T_h-T_l}{L}=\sigma (T_h-T_l)\,,
\end{equation}
consistently with Eq. \eqref{steady}.


Let us focus now  on the second possibility, namely the anisotropically coupled  $XY$-model with $L$ interacting spins $\mathbf{S}_i$. The corresponding Hamiltonian  is given by
\begin{equation}
    \mathcal{H}_{XY}^{\,a}=\sum_{i=1}^L\frac{\ell_i^2}{2}-\mathcal{J}_x\sum_{\langle i,j\rangle } S_i^xS_j^x-\mathcal{J}_y\sum_{\langle i,j\rangle}S_i^y S_i^y\,.
    \label{H9}
\end{equation}

\begin{figure}
\centering
\includegraphics[width=4.2cm, angle=90]{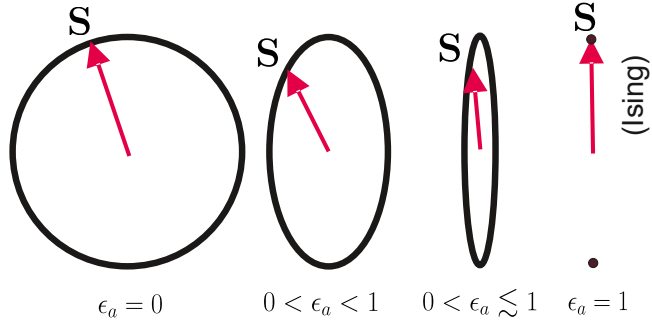}
  \caption{Schematic representation of the anisotropic $XY$ coupling.}
  \label{SQUEM}
\end{figure}

We define $\mathcal{J}_x\equiv \mathcal{J}(1+\epsilon_a)/2$ and $\mathcal{J}_y\equiv \mathcal{J}(1-\epsilon_a)/2$ with $|\epsilon_a|\leq 1$ and  $\mathcal{J}>0$ . This Hamiltonian can be rewritten in polar coordinates as follows:
\begin{equation}
\label{aniso}
    \mathcal{H}_{XY}^{\,a}= \sum_i^L \frac{p_i^2}{2}+\frac{1}{2}\sum_{\langle i,j \rangle}[1+\epsilon_a-\cos(\theta_i-\theta_j)-\epsilon_a\cos(\theta_i+\theta_j)]
\end{equation}
Without loss of generality, we set moment of inertia  and  exchange interaction $\mathcal{J}$ equal to unity. Notice that $\theta_i=0\,\,,\forall i$,  leads to zero potential energy, $\forall \epsilon_a$. Notice also that $\epsilon_a=\pm 1$  correspond to the Ising model along the $y$ and $x$ axes respectively, whereas  $\epsilon_a=0$ recovers the standard isotropic $XY$-model (see Fig.~\ref{SQUEM}). The equations of motion are the same as in  Eq.~\eqref{langevin}, the forces being now written as follows:
\begin{equation}
    F_i=-\sum_{\langle j \rangle} \left[\sin(\theta_i-\theta_j)+\epsilon_a\sin(\theta_i+\theta_j)\right]\,.
\end{equation}
The heat flux of the anisotropically coupled $XY$ model is given by
\begin{equation}
    J_i=\frac{p_i+p_{i+1}}{2}\sin(\theta_i-\theta_{i+1})+\epsilon_a\frac{p_i-p_{i+1}}{2}\sin(\theta_i+\theta_{i+1})\,.
    \label{flux2}
\end{equation}


 For both models, we implement the equations of motion with Velocity Verlet algorithm \cite{Verlet1967, PaterliniFerguson1998}. In our simulation, we set the step size $dt=0.01$, $\gamma_h=\gamma_l=1.0$, and the temperatures $T_{h/l}=T(1\pm \Delta)$, with $\Delta\equiv 0.125$ in such way that $T=(T_h+T_l)/2$. The coordinates and momenta are initially set to zero, naturally conserving the sum of the $L$ momenta. 

 We  directly compute the averages from Eqs. \eqref{flux1} and \eqref{flux2},  assuming a transient time equal to $10^{10}$, and averaging the heat flux along $4\times 10^8$ time-steps with 200 experiments. After that, Eq.~\eqref{steady} is used for 23  different values for the temperature $T$, with   $\epsilon_l$ and $\epsilon_a$ from 0 to $0.7$ by steps of 0.1.

 \section{Results}

\begin{figure}[htb]
    \centering
    \includegraphics[width=5.2cm]{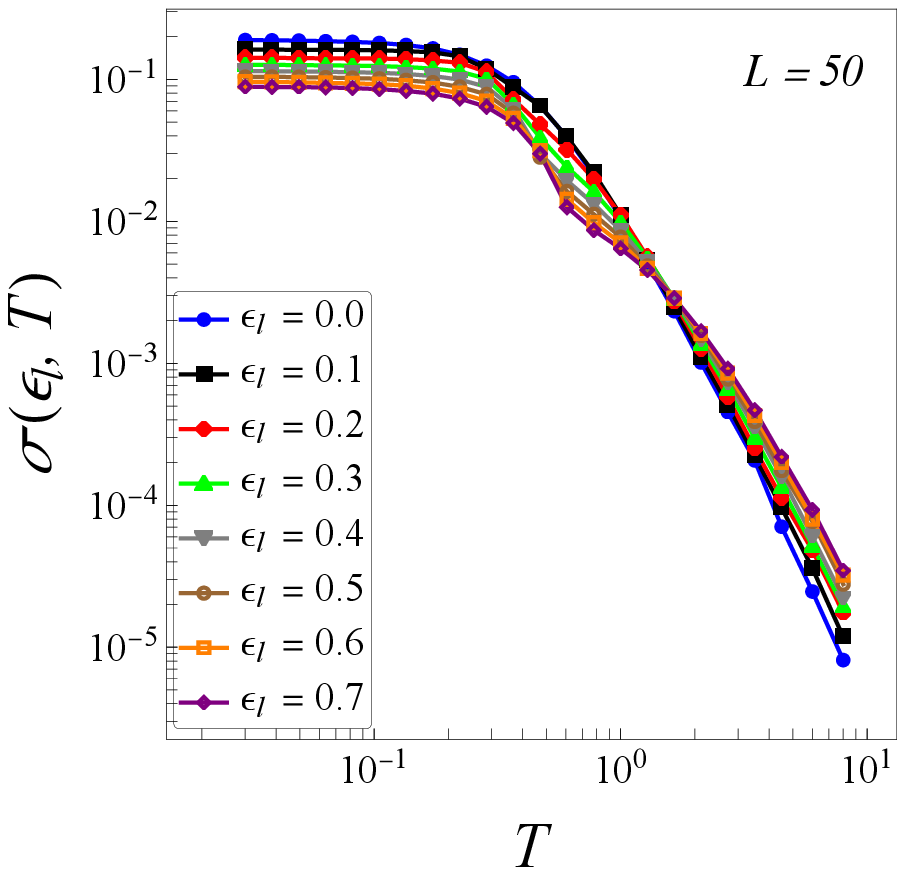}
    \includegraphics[width=5.0cm]{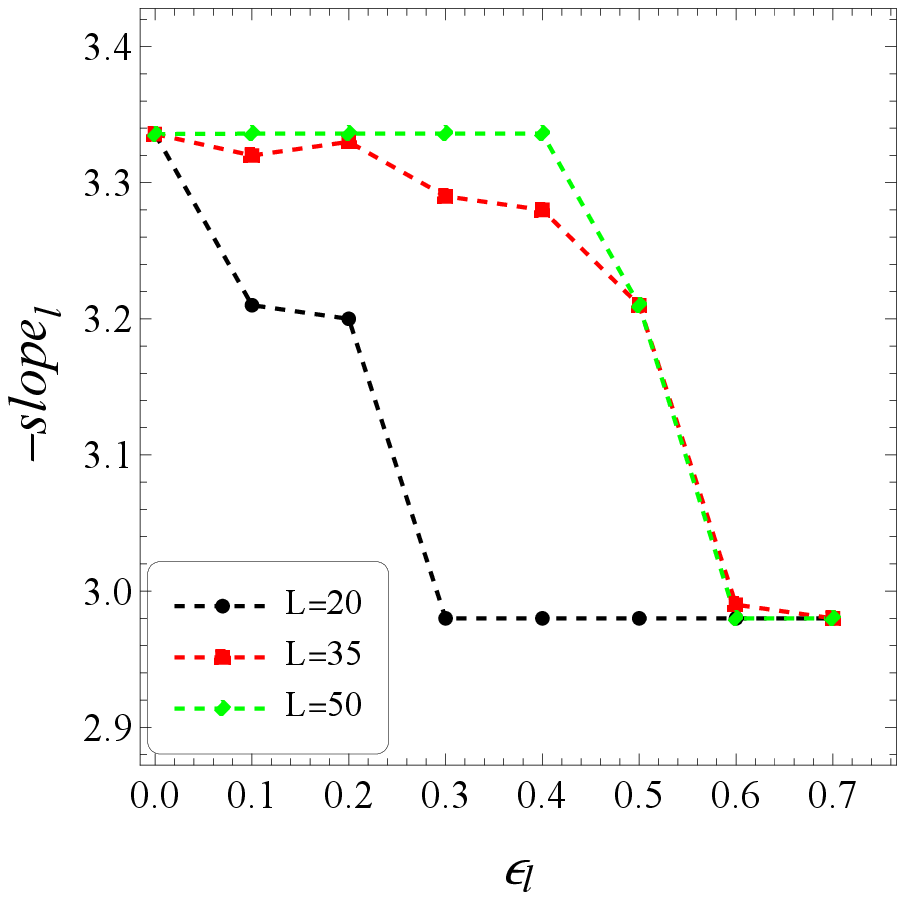}
    \includegraphics[width=5.2cm]{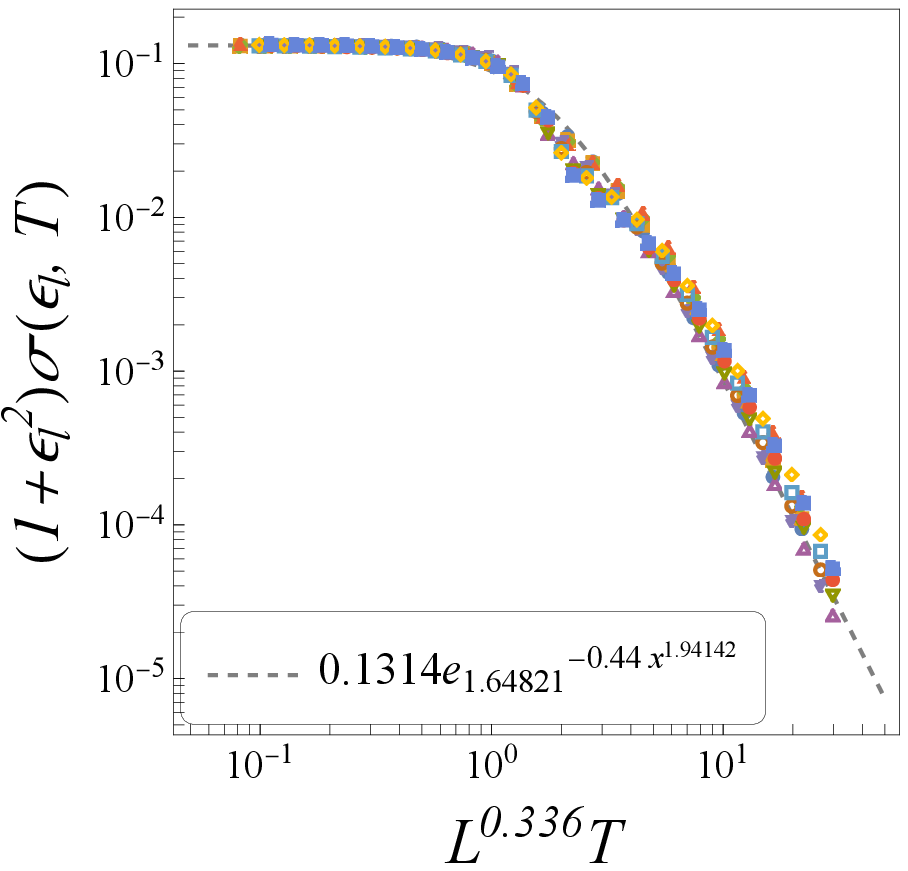}
    \caption{{\it Left:} Thermal conductance of the first anisotropic model as a function of temperature for one-dimensional lattice structure and the local coupling constant for $L=50$. {\it Center:} Plot of -slope versus $\epsilon_l$ for $L=20,35,50$. All the curves approach the same saturation  value $slope_l\simeq -3.0$. {\it Right:} Collapse with a stretched $q$-exponential form, from $\epsilon_l=0.4$ to $\epsilon_l=0.7$ with $L=20,35,50$. }
    \label{1st}
\end{figure}

\begin{figure*}[htb]
    \centering
    \includegraphics[width=5.2cm]{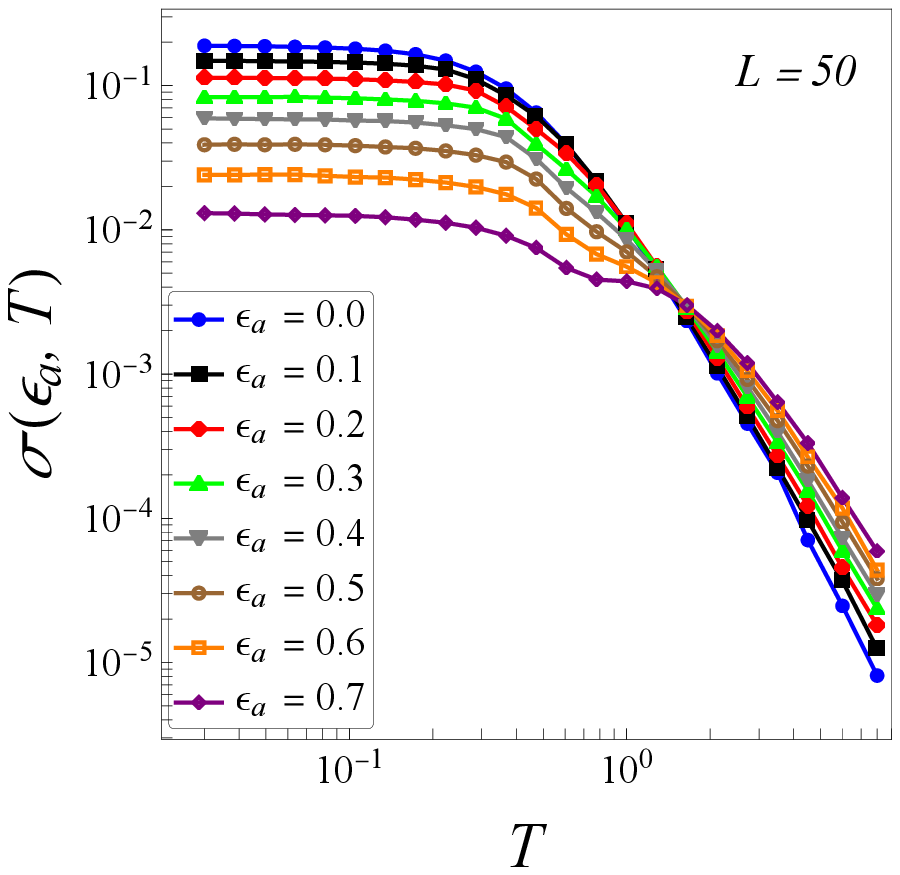}
    \includegraphics[width=5.0cm]{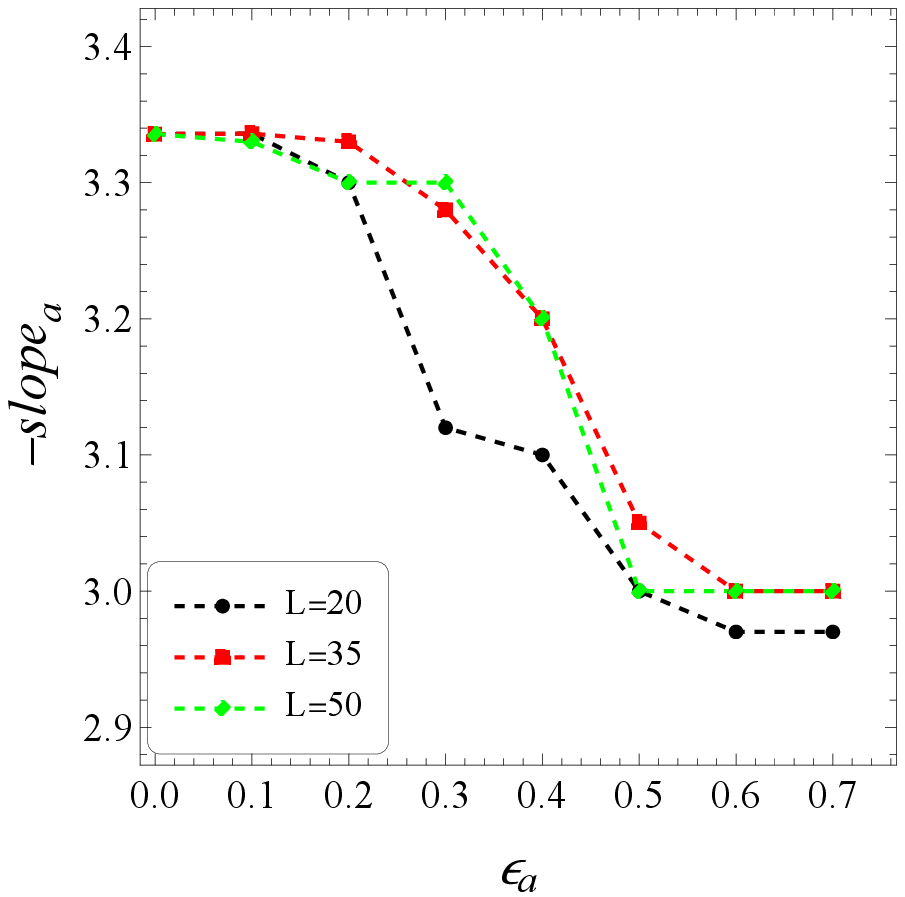}
    \includegraphics[width=5.2cm]{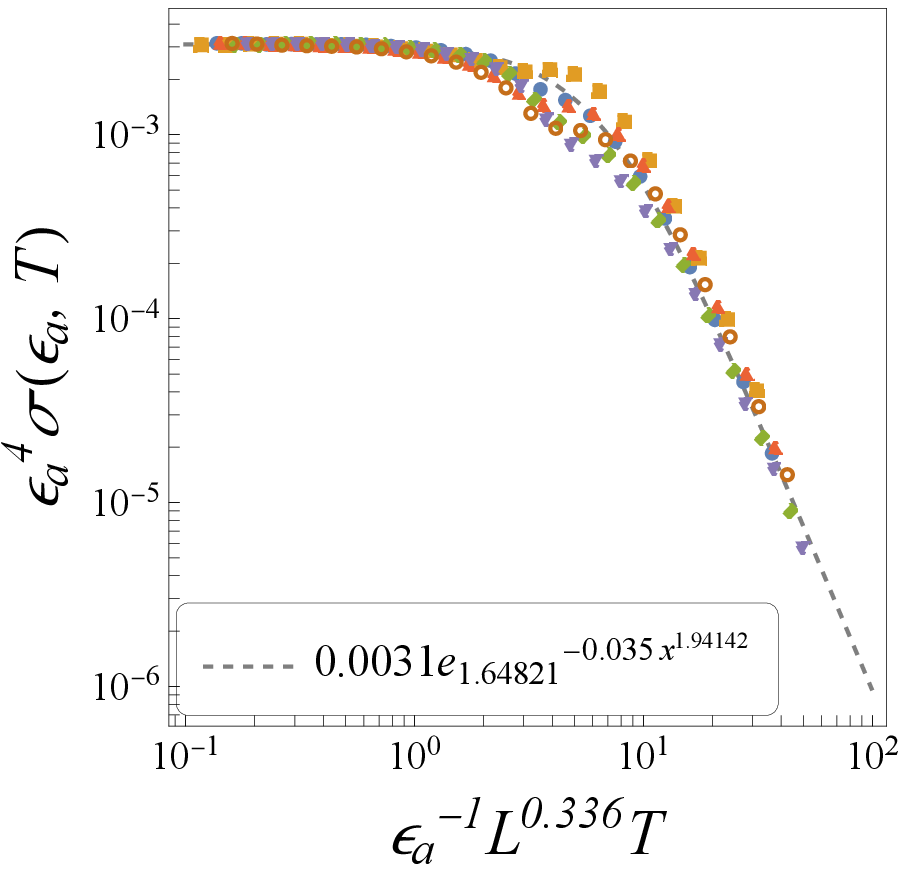}
        \caption{{\it Left:} Thermal conductance of the second anisotropic model as a function of temperature for one-dimensional lattice structure and the local coupling constant for $L=50$. {\it Center:} Plot of  -slope versus $\epsilon_a$ for $L=20,35,50$. All the curves approach the same saturation  value $slope_a\simeq -3.0$. {\it Right:} Collapse with a stretched $q$-exponential form, where $(A,)$,  for $\epsilon_a=0.6$ and $\epsilon=0.7$  with $L=20,35,50$. }
    \label{2nd}
\end{figure*}

We observe in Figs. \ref{1st} and \ref{2nd} that, at low temperatures, the thermal conductance $\sigma$ decreases for increasing anisotropic parameters $\epsilon_l$ and $\epsilon_a$: $\sigma$ decreases for the first model (Fig. \ref{1st}) slower than for the second one (Fig. \ref{2nd}).  
For the first model, for instance, the decrease is related to the fact that, at small oscillations ($\theta_i \simeq 0$), an additional force $-2\epsilon_l \theta_i$ emerges
which reduces  the mean heat flux, hence the thermal conductance. A similar effect is present in the second model with regard to $\epsilon_a$.

At intermediate temperatures a  crossover  becomes preliminary evident. This is due to the fact that the rotators are now at excited states, and therefore start being angularly constrained because of the anisotropy, as depicted in Fig.~\ref{SQUEM}.
We can see  in Fig.~\ref{1st} that, after that intermediate regime, the  absolute value of the slope  reduces more and more until it saturates, making the cases $\epsilon_l=0.5,0.6,0.7$  to virtually coincide. The Ising regime starts appearing in this limit, the potential energy per particle   asymptotically becoming   $\frac{1}{2}\sum_{\langle j\rangle}[1-S_i^y S_j^y]+\epsilon_l (S_i^x)^2$.

We can collapse all thermal conductances of both models, except in the crossover region, with a stretched $q$-exponential Ansatz (see \cite{Pickupetal2009,TsallisLima2023}), defined as 
\begin{equation}
y(x)=e_q^{-Bx^\eta}\;\;\;(x \ge 0, q\ge 1, \eta > 0, B>0) \,,
\label{moregeneral}
\end{equation}
where $e_q^z\equiv[1+(1-q)z]^\frac{1}{1-q}\,(e_1^z=e^z)$. Consistently with this Ansatz, we verify that, in the thermodynamic limit ($L\gg 1 $), $\sigma(\epsilon_l,T)\propto \sigma(\epsilon_a,T)\propto T^{-\frac{\eta}{q-1}}$, where $(\eta,q)\approx (1.94,1.65)$  thus yielding the slope $\eta/(q-1)\approx 3.0$. This value is already shown in Figs.~\ref{1st} (center) and \ref{2nd} (center). Another important observation is that $\kappa=\sigma L$ does not depend on the system size $L$. This is a simple consequence from the fact that, at fixed temperatures, we have $\sigma\sim L^{-\gamma\frac{\eta}{q-1}}$, with $\gamma=0.336$, hence $\sigma \propto L^{-1}$, thus validating, through both anisotropic models, the Fourier's law in the Ising limit.


Let us emphasize that, in the isotropic $XY$ model, $\sigma_{XY}\sim  T^{-3.34}$ ~\cite{TsallisLima2023} while, in the Ising limit, we have $\sigma_{Ising} \sim T^{-3.0}$. The parameters $(q,\eta,\gamma)$ of the XY and Ising models are sensibly different. However, when all those parameters are combined together, a remarkable numerical result is obtained, namely that the thermal conductivity $\kappa$ becomes asymptotically independent of the lattice size, thus  obeying Fourier's law. 
It should be noted that $\frac{\eta\gamma}{q-1}\approx 1$ for both the Ising and $XY$ linear chains. It is in fact plausible to expect that, for the $n$-vector models, $\frac{\eta\gamma}{q-1}\approx 1$ for all values of $n$ .

\section{Final remarks}
Let us summarize that both anisotropic models studied here have one and the same temperature-exponent for the thermal conductivity at high temperatures, and differ from that corresponding to the isotropic $XY$ model. To be more precise, the behavior in the regions  $\epsilon_l>1/2$ and $0.5<\epsilon_a\approx 1$ suggests that the thermal conductivity of the Ising model decays as $T^{-3}$ law. Moreover, we have verified that the stretched $q$-exponential Ansatz provides a satisfactory description over all temperatures. 
The present first-principle numerical approaches of momentum-conserving ferromagnetic systems neatly help understanding their thermal transport properties, and ultimately validate Fourier's law, phenomenologically proposed two centuries ago.

We acknowledge fruitful discussions with U. Tirnakli and D. Eroglu, useful remarks from U. B. de Almeida and L. M. D. Mendes, as well as  partial financial support by CNPq and FAPERJ (Brazilian agencies). We also acknowledge LNCC (Brazil) for allowing us to use the Santos Dumont (SDumont) supercomputer.

\end{document}